\documentclass[aps,prx,a4paper,twocolumn,superscriptaddress,citeautoscript]{revtex4-1}
\usepackage{color}
\usepackage[english]{babel}
\usepackage{lmodern}
\usepackage[T1]{fontenc}
\usepackage{amsmath}
\usepackage{units}
\usepackage{grffile}
\usepackage{color}
\usepackage[bottom]{footmisc}
\usepackage{booktabs}
\usepackage{dcolumn}
\usepackage{graphicx}
\graphicspath{{figs/}}

\newcommand\beq{\begin{equation}}
\newcommand\eeq{\end{equation}}

\newcommand\vk{\mathbf{k}}
\newcommand\ek{\varepsilon_{\mathbf{k}}}

\newcommand\hn{\hat{n}}

\newcommand\hN{\hat{N}}
\newcommand{\vS}{\vec{S}}
\newcommand{\vL}{\vec{L}}

\newcommand{\scsc}{IHSC\,}
\newcommand{\Jinv}{J_{\mathrm{inv}}\,}
\newcommand{\full}{A$_{3}$C$_{60}$\,}
\newcommand{\fullK}{K$_{3}$C$_{60}$\,}
\def\t1u{$t_{1u}$}
\newcommand{\hcf}{h_{\mathrm{CF}}}

\begin{document}
%
\title{Enhancing superconductivity in A$_3$C$_{60}$ fullerides}
\author{Minjae Kim}
\email{garix.minjae.kim@gmail.com}
\affiliation{Centre de Physique Th\'eorique, \'Ecole Polytechnique, CNRS, Universit\'e Paris-Saclay, 91128 Palaiseau, France}
\affiliation{Coll\`ege de France, 11 place Marcelin Berthelot, 75005 Paris, France}
\author{Yusuke Nomura}
\affiliation{Centre de Physique Th\'eorique, \'Ecole Polytechnique, CNRS, Universit\'e Paris-Saclay, 91128 Palaiseau, France}
\author{Michel Ferrero}
\affiliation{Centre de Physique Th\'eorique, \'Ecole Polytechnique, CNRS, Universit\'e Paris-Saclay, 91128 Palaiseau, France}
\affiliation{Coll\`ege de France, 11 place Marcelin Berthelot, 75005 Paris, France}
\author{Priyanka Seth}
\affiliation{Institut de Physique Th\'eorique (IPhT), CEA, CNRS, 91191 Gif-sur-Yvette, France}
\affiliation{Coll\`ege de France, 11 place Marcelin Berthelot, 75005 Paris, France}
\author{Olivier Parcollet}
\affiliation{Institut de Physique Th\'eorique (IPhT), CEA, CNRS, 91191 Gif-sur-Yvette, France}
\affiliation{Coll\`ege de France, 11 place Marcelin Berthelot, 75005 Paris, France}
\author{Antoine Georges}
\affiliation{Coll\`ege de France, 11 place Marcelin Berthelot, 75005 Paris, France}
\affiliation{Centre de Physique Th\'eorique, \'Ecole Polytechnique, CNRS, Universit\'e Paris-Saclay, 91128 Palaiseau, France}
\affiliation{Department of Quantum Matter Physics, University of Geneva, 24 Quai Ernest-Ansermet, 1211 Geneva 4, Switzerland}

\date{\today}
\begin{abstract}
Motivated by the recent experimental report of a possible light-induced superconductivity in \fullK at high temperature
[Mitrano {\it et al.}, Nature 530, 451 (2016)],
we investigate theoretical mechanisms for enhanced superconductivity in \full fullerenes.
We find that an `interaction imbalance' corresponding to a smaller value of the
Coulomb matrix element for two of the molecular orbitals in comparison to the third one,
efficiently enhances superconductivity.
Furthermore, we perform first-principle calculations of the changes in the electronic structure and
in the screened Coulomb matrix elements of \fullK, brought in by the deformation associated with
the pumped $T_{1u}$ intra-molecular mode. We find that an interaction imbalance is indeed induced,
with a favorable sign and magnitude for superconductivity enhancement.
The physical mechanism responsible for this enhancement consists in a stabilisation of the intra-molecular states
containing a singlet pair, while preserving the orbital fluctuations allowing for a coherent inter-orbital delocalization of the pair.
Other perturbations have also been considered and found to be detrimental to superconductivity.
The light-induced deformation and ensuing interaction imbalance is shown to bring superconductivity further into the strong-coupling regime.
\end{abstract}
\maketitle
\section{Introduction}
\label{sec:intro}

Alkali-doped fullerenes \full (A = K, Rb, Cs)
are a remarkable family of materials, which have the highest superconducting (SC) transition temperature
among molecular superconductors ($T_{c}\sim$40K)~\cite{Hebbard_Nature_1991,palstra_ssc_1995}.
Even though the gap symmetry is $s$-wave, the mechanism of SC has been the subject of much debate.~\cite{capone_rmp_2009,Gunnarsson_book_2004,han_A3C60_prl,VARMA_Science_1991,Schluter_PRL_1992,Mazin_PRB_1992,
Zhang_PRL_1991,Rice_PRB_1991,Asai_PRB_1992,
Suzuki_JPSJ_2000,Takada_1993,CHAKRAVARTY_Science_1991,
Jiang_PRB_2016,Capone_PRL_2001,capone_scsc_science_2002,
Capone_PRL_2004,nomurae_A3C60_sciadv,
Nomura_review_2016,
Koga_PRB_2015,Steiner_orb_frez_2016}.
Indeed, the narrow bandwidth ($W\sim$0.5 eV), the intramolecular Coulomb interaction ($U\sim$0.6 eV), and
the typical frequency of the relevant intramolecular Jahn-Teller phonons
($\omega_{\rm ph}\sim$0.1 eV)
are comparable energy scales, which raises questions about the validity of
conventional phonon-mediated SC mechanisms~\cite{ramirez_s_wave_1992,gunnarsson_A3C60_rmp,capone_rmp_2009,han_A3C60_prl}.
%
The experimental observation ~\cite{Ganin_Nature_Mat_2008,Takabayashi_Science_2009,Ganin_Nature_2010,
Klupp_DJT_Nature_comm_2012,Kasahara_PRB_2014,
Potocnik_sci_rep_2014,zadik_JTM_2015,
Ihara_Cs3C60_NMR_PRB_2010,Ihara_Cs3C60_NMR_EPL_2011,Wzietek_Cs3C60_gap_PRL_2014,
Kawasaki_JPSJ_2013} of a Mott-insulating phase close to the SC phase in Cs$_3$C$_{60}$ emphasizes the
importance of strong electronic correlations for these materials  and of their interplay with SC.~\cite{Lof_PRL_1992,Chakravarty_PRL_1992,Gunnarsson_PRL_1992}

Recently, a remarkable experiment~\cite{mitrano_K3C60_2016}  by Mitrano \textit{et al.} reported a large enhancement in carrier mobility
and the opening of an optical gap when exciting \fullK with a mid-infrared femto-second light-pulse
in the frequency range $80-200$~meV ($19-48$~THz).
These observations were interpreted as the signature of a light-induced SC at temperatures (up to $T\sim 100$~K) much higher
than the equilibrium $T_c\sim 20$~K.
This experiment raises a number of intriguing questions. On the theory side, one of the most pressing ones is whether there
are indeed perturbations of the system which could lead to enhanced SC. And if so, whether such perturbations
are likely to be induced by the mid-infrared excitation of Ref.~\onlinecite{mitrano_K3C60_2016}. Providing some answers
to these two questions is the main purpose of this article.

In order to address these questions, we shall place ourselves within the theoretical framework reviewed
in Ref.~\onlinecite{capone_rmp_2009}, which is one of the most successful ones at explaining
the physical properties and SC of fullerenes. This approach focuses on the set of three near-Fermi level $t_{1u}$ electronic states derived from
the lowest unoccupied molecular orbitals. It furthermore recognizes two important ingredients for the physics of these
materials.
First, as pointed out early-on ~\cite{gunnarsson_A3C60_rmp}, the
intra-molecular Jahn-Teller phonons with $H_g$ symmetry play
a key role in the pairing. This leads to a reversal of the sign of the effective intra-molecular exchange
acting on the $t_{1u}$ states (inverted Hund's coupling)~\cite{Capone_PRL_2001,Capone_PRB_2000}.
Second, the intra-molecular repulsion (Hubbard $U$)
is comparable or larger than the bandwidth. This was shown~\cite{capone_scsc_science_2002} by Capone \textit{et al.} to lead to a strong-coupling regime
of the SC induced by the inverted Hund's coupling, near the Mott transition.
This `inverted Hund's coupling' model~\cite{capone_rmp_2009} for the SC of fullerenes (\scsc ) is introduced in more details in Sec.~\ref{sec:model}.
Recently, an extensive first-principles study~\cite{nomurae_A3C60_sciadv} by Nomura \textit{et al.,}  has been successful
at reproducing many properties of these materials and also validated the qualitative premises of the \scsc model.

The main finding of this article is that a perturbation which efficiently enhances SC indeed exists. This perturbation
consists of an 'interaction imbalance', in which the intra-molecular Coulomb repulsion is larger by an amount $dU>0$
in one of the $t_{1u}$ orbitals, as compared to the two others.
For example, at $dU/U\simeq 0.2$, the SC gap, order parameter, and critical temperature are increased by
large amounts (a factor of $3.5$, $1.6$, and $1.8$, respectively).
We furthermore demonstrate, using first principle calculations, that the THz pumping of the mid-infrared $T_{1u}$ structural
mode considered in Ref.\onlinecite{mitrano_K3C60_2016} indeed leads to such a perturbation with a favorable sign $dU>0$,
and estimate the amplitude of the corresponding SC enhancement.
Such an interaction imbalance was indeed discussed in Ref.~\onlinecite{mitrano_K3C60_2016}, in relation
also to previous work in which modulation of $U$ by light-excitation of organic materials was demonstrated.\cite{singla_molU_2015}
However, the effect of this perturbation on SC
was not investigated theoretically.

Obviously, the theoretical analysis presented here is an equilibrium one, while the observation
made in Ref.\onlinecite{mitrano_K3C60_2016} results from an out-of-equilibrium pump-probe experiment.
As such, our work is a first step towards a full understanding of the experimental results.
The main message
is that the light-induced perturbation could indeed drive
the system into a more strongly-coupled SC regime.

\section{Inverted-$J$ coupling model and symmetry-breaking perturbations}
\label{sec:model}

\subsection{Inverted-Hund model and strongly-correlated superconductivity}

Let us briefly recall the key ingredients entering the \scsc
model for the superconductivity of \full fullerenes, which
will be used throughout this article (see Ref.~\onlinecite{capone_rmp_2009} for a review). The model focuses on
the three bands originating from the molecular \t1u states. The intra-molecular Coulomb repulsion $U$ associated with
these states is comparable to their bandwidth $U/W \gtrsim 1$, so that the system is in the regime of strong correlations
(consistent with the proximity of a Mott insulating state).

The $H_g$ Jahn-Teller (JT) modes play an important role in the pairing and superconductivity.
Their frequency is a significant
fraction of the bandwidth ($\omega_{\mathrm{ph}}/W\sim 0.2$).
Because of this, and of strong correlations, conventional
Migdal-Eliashberg theory does not apply.\cite{grimaldi_nonadiabatic} In the simplest version of the model, the phonon degrees of freedom are integrated
out, leaving a purely electronic model for the three \t1u states. The distinctive feature of this model is that the effective Hund's coupling $\Jinv$
applying to these states has an inverted (antiferromagnetic) sign, in contrast to the conventional purely electronic
Hund's coupling: $\Jinv=J_{\rm H}+J_{\rm JT} < 0$.
The local (intra-molecular) interactions applying to the \t1u states thus take the conventional Kanamori form:
\begin{eqnarray} \label{eq:Kanamori}
H_{\rm int}=(U-3\Jinv)\frac{\hN(\hN-1)}{2}-2\Jinv\vS^{2}-\frac{\Jinv}{2}\vL^2
\end{eqnarray}
to which one should of course add the kinetic energy term (inter-molecular hopping):
\begin{equation}
\label{eq:kinetic}
H_0 = \sum_{\vk\sigma\alpha}\,\ek  d^\dagger_{\vk\sigma\alpha}d_{\vk\sigma\alpha}
\end{equation}
Here, $\vS$ and $\vL$ are the spin and orbital angular  momentum operators associated with
the three orbitals, labelled by $\alpha$ (see e.g. Ref.\onlinecite{georges_Hund_review_annrev_2013}).

For future reference, we note that the Kanamori interaction (Eq.\ref{eq:Kanamori}) can be rewritten\cite{georges_Hund_review_annrev_2013} as
$H_{\rm int}=H_{nn}+H_{\rm sf}+H_{\rm ph}$ in which $H_{nn}$
involves only density-density interactions, $H_{\rm sf}$ is a spin-flip term and the `pair-hopping' term $H_{\rm ph}$.
$H_{\rm ph}$ transfers a singlet pair from one orbital to another one, namely:
\begin{eqnarray} \label{eq:Kanamori_dec}
&H_{nn}&\,=\,U \sum_{\alpha}\,\hn_{\alpha,\uparrow}\hn_{\alpha,\downarrow} \nonumber\\
&+& (U-2\Jinv) \sum_{\alpha\ne\beta}\,\hn_{\alpha,\uparrow}\hn_{\beta,\downarrow} \nonumber\\
&+& (U-3\Jinv) \sum_{\alpha<\beta,\sigma}\,\hn_{\alpha,\sigma}\hn_{\beta,\sigma} \nonumber\\
&H_{\rm sf}&= -\Jinv \sum_{\alpha\ne\beta} d^\dagger_{\alpha,\uparrow}d_{\alpha,\downarrow}d^\dagger_{\beta,\downarrow}d_{\beta,\uparrow} \nonumber\\
&H_{\rm ph}&= \Jinv \sum_{\alpha\ne\beta} d^\dagger_{\alpha,\uparrow}d^\dagger_{\alpha,\downarrow}d_{\beta,\downarrow}d_{\beta,\uparrow}
\end{eqnarray}

The rationale for $\Jinv < 0$ is that the JT modes dynamically lift the degeneracy of the three states, hence
favoring the $S=1/2, L=1$ local configuration when three electrons occupy the \t1u molecular orbital
(half-filled shell, as appropriate for \full). This is in contrast to the $S=3/2,L=0$ configuration favored by the electronic Hund's coupling.
The six-fold degenerate $S=1/2, L=1$ configuration basically consists in one singlet pair occupying any of the three
orbitals, and a third lone electron in one of the two remaining ones (see Sec.~\ref{sec:multiplets}).
Through the intra-molecular pair-hopping term $H_{ph}$,
this pair is delocalised between any of the three orbitals, and these orbital fluctuations promote SC~\cite{nomurae_A3C60_sciadv,Steiner_orb_frez_2016}.

The basic reason for which SC is driven into the strong-coupling regime (hence leading to high $T_c$) in this model is
the following~\cite{capone_scsc_science_2002,capone_rmp_2009}.
As the Mott transition is approached, charge fluctuations are suppressed and the quasiparticle bandwidth
is renormalized by the Brinkman-Rice phenomenon~\cite{brinkman_rice_mit_prb_1970}
down to $Z W$ with $Z$ the quasiparticle spectral weight.
In contrast, the  $\Jinv$ interaction does not couple to the charge sector, but only to the spin and orbital sectors, and hence this
interaction is basically unrenormalized. As a result, the effective attractive coupling $\Jinv/ZW$ becomes large and the SC enters the strong-coupling regime.
As $U/W$ is increased from the weak-coupling regime, the SC gap, $T_c$ and the SC order parameter increase.
The latter decreases again as the Mott transition is approached, following a dome-like behavior (see Fig.~\ref{fig:degenerate case} in
Appendix).

Simplified as it may be, several observations provide a justification to the \scsc model as a minimal model for
the SC of \full fullerenes.
The spin gap corresponding to the transition between low-spin $S=1/2$,$L=1$ and
high-spin $S=3/2$,$L=0$ configurations ($\sim 5\Jinv\sim0.1$ eV, see Sec.\ref{sec:multiplets}) is indeed observed
in nuclear magnetic resonance experiments~\cite{prassides_NMR_1999,brouet_NMR_2002}.
Also, a recent first-principle investigation\cite{nomurae_A3C60_sciadv} treating electronic correlations according to (\ref{eq:Kanamori}) but
explicitly taking into account JT phonon modes turns out to
quantitatively describe the phase diagram of \full materials as a function of unit-cell volume.

\subsection{Symmetry-breaking perturbations: $U$-imbalance and crystal-field splitting}

Our strategy in this article is to consider perturbations of the Hamiltonian (\ref{eq:Kanamori},\ref{eq:kinetic})
and to identify which perturbations can lead to an enhancement of superconductivity.
We consider specifically two kinds of perturbations, of which both break the symmetry between \t1u orbitals.

The first one (`$U$-imbalance') is a perturbation that modifies the intra-molecular Coulomb energy $U$ for two of the orbitals
(conventionally chosen to be $x$ and $y$) as compared to the third one:
\begin{equation} \label{eq:AsyU}
H_{dU}=-dU(\hn_{x,\uparrow}\hn_{x,\downarrow}+\hn_{y,\uparrow}\hn_{y,\downarrow})
\end{equation}
For $dU>0$, the Coulomb energy becomes smaller for these two orbitals, while it becomes larger for $dU<0$.
As we shall see, these two signs of the perturbation lead to very different effects on SC.

The second perturbation (`crystal-field') is a splitting between the molecular energy levels of the $x,y$ orbitals
and that of the $z$-orbital, hence lifting the degeneracy of the \t1u triplet at the one-electron level:
\begin{equation} \label{eq:CF}
H_{\rm CF}=\hcf\,(\hn_{x}+\hn_{y})
\end{equation}
In contrast to the $U$-imbalance case, this commutes with $L_z$: $[H_{\rm CF},L_{z}]=0$.
This perturbation was also considered recently in Ref.~\onlinecite{Hoshino_PRB_2016}.
As shown there (see also Appendix.\ref{sec:Negative crystal field}),
$\hcf>0$ and $\hcf<0$ are equivalent in the presence of particle-hole symmetry.

In Sec.~\ref{sec:Tera-Hertz excitation}, we motivate the study of these two perturbations above from
first-principle studies of \fullK subject to a THz pump exciting the $T_{1u}$ structural mode. The
reader is thus directed to this section for physical motivations of these perturbations.

\subsection{Methods and Observables}
\label{sec:methods}

We solve the \scsc model in the presence of the perturbations above using dynamical mean-field theory (DMFT, for a review see e.g. Ref.\onlinecite{DMFT_rmp_1996}).
For simplicity, the non-interacting density of states (DOS) associated with the dispersion $\ek$ is chosen
to be a semi-circular with width $W$.
This corresponds to a Bethe lattice in the limit of infinite coordination, for which DMFT becomes exact.

In order to solve the model in the superconducting phase, we introduce a Nambu spinor
$\Psi^{\dagger}_{\alpha} \equiv (d_{\alpha\uparrow}^{\dagger},d_{\alpha\downarrow}^{})$
for each orbital $\alpha=x,y,z$. For the chosen semi-circular DOS,
the dynamical mean-field ${\cal G}$ (Weiss function)\cite{DMFT_rmp_1996}
entering the impurity model is then given by the self-consistency condition:
\begin{equation} \label{eq:NambuSCS}
\hat{\cal{G}}^{-1}_{\alpha}(i\omega_{n}) = i\omega_{n}\textbf{1} + \mu\tau_{z} - (\frac{W}{4})^{2}\tau_{z}\hat{G}_{\alpha}(i\omega_{n})\tau_{z}
\end{equation}
In this expression, we have used Nambu matrix notations (with $\tau_z$ the Pauli matrix), $\hat{G}_{\alpha}(\tau) \equiv - \langle T \Psi_{\alpha}^{} (\tau) \Psi_{\alpha}^{\dagger}(0) \rangle$
denotes the local (on-site) Green's function in matrix form, and the $\omega_n$ are Matsubara frequencies.
The chemical potential $\mu$ is adjusted to insure an occupancy of  three electrons per site.

We solved the DMFT quantum impurity problem using the hybridisation expansion continuous-time quantum Monte Carlo (CTQMC) solver\cite{gull_CTQMC_rmp_2011},
as implemented in the TRIQS library\cite{TRIQS_Package_2015,TRIQS_CTQMC_2016}. It should be noted that, in order to
properly sample the SC phase, four-operators insertions are essential in Monte-Carlo updates.\cite{Karim_CT-HYB_2015,Semon_CT-HYB_SC_2014}

In order to characterize the SC phase, we focus in particular on the following observables.
\begin{itemize}
\item The orbitally-resolved pairing amplitude $P_{\rm sc}^{\alpha}$ and total SC order parameter  $P_{\rm sc}$ (summed over all orbitals),
obtained from the off-diagonal component of the Green's function in Nambu matrix form:
\beq
P_{\rm sc}^\alpha = G^{12}_{\alpha}(\tau=0^{+})\,\,\,,\,\,\,
P_{\rm sc}=\sum_{\alpha=x,y,z}\,G^{12}_{\alpha}(\tau=0^{+})
\eeq
\item The quasiparticle spectral weight for each orbital, $Z_\alpha$, defined from the
diagonal component of the retarded self-energy:
\beq
Z_{\alpha} =\left[1-\frac{\partial\mathrm{Re}\Sigma_{{\rm ret},\alpha}^{11}}{\partial\omega}|_{\omega=0}\right]^{-1}
\eeq
In practice, this can be accurately approximated at low temperature from the first Matsubara frequency:
\beq
Z_{\alpha} \,\simeq\,\left[1-\frac{Im\Sigma^{11}_{\alpha}(i\omega_{0})}{\omega_{0}}\right]^{-1}
\eeq
\item From the low-frequency expansion of the Nambu Green's function, we see that the dispersion of the $\alpha$-orbital
quasiparticles in the SC state read : $\omega^\alpha_\pm(\vk)=\pm Z_\alpha \sqrt{(\ek-\mu)^2+(\Sigma^{12}_{\alpha}(0))^2}$.
Hence the SC gap for orbital $\alpha$ is given by:
\beq
\Delta_{\alpha}\,=\,Z_{\alpha}\Sigma^{12}_{\alpha}(0)\simeq Z_{\alpha}\Sigma^{12}_{\alpha}(i\omega_0)
\eeq
\item The SC critical temperature $T_{c}$ was estimated by stabilizing a SC solution for several different temperatures,
and fitting the corresponding temperature-dependence of the order parameter to a mean field-form $P_{\rm sc}(T) = C \sqrt{T_c-T}$
for three temperatures closest to the boundary of the SC phase.

\end{itemize}

The parameters in all the following will be chosen in accordance with values determined in previous work\cite{nomura_C60_cRPA_2012,nomura_C60_phonon_2015}
as appropriate for the description of \fullK. Unless stated otherwise, we take $U=W$ and $\Jinv = - 0.04\,W$. Most of the results
below are displayed at the lowest temperature we have studied in the SC phase, namely $T= 0.005 W$.
For K$_{3}$C$_{60}$, the $t_{1u}$ bandwidth is of order  $W\simeq 0.50$ eV and hence $U=0.50$~eV, $\Jinv = -20$~meV.

\section{Results}
\label{sec:results}

In this section, we present our main findings regarding the effect of the above perturbations on superconductivity.

\subsection{Imbalanced Coulomb interaction}

\begin{figure}[t]
\includegraphics[width=\columnwidth]{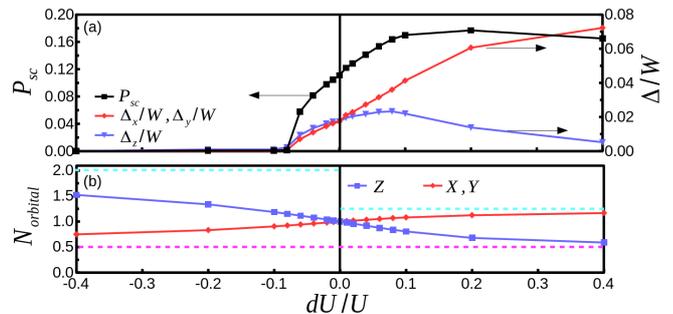}
\caption{(Color)
(a) Total superconducting (SC) order parameter $P_{\rm sc}$ and orbitally-resolved SC gaps $\Delta_{\alpha}/W$
as a function of the interaction imbalance $dU/U$ (as defined in Eq.\ref{eq:AsyU}).
(b) Orbital occupancies as a function of $dU/U$.
Dashed lines indicate asymptotic values of the orbital occupancies discussed in the text,
for $dU<0$ and $dU>0$, respectively.
\label{fig:imbalance_results}
}
\end{figure}

Fig.\ref{fig:imbalance_results} (a) displays the superconducting order parameter and orbitally-resolved gaps as a function of the
imbalance $dU/U$.
A negative imbalance $dU<0$ (i.e. having a larger Coulomb repulsion on two orbitals as compared to the third one) is detrimental to SC.
In contrast, a positive imbalance (smaller repulsion on two of the orbitals) leads to a remarkable enhancement of superconductivity.
The largest enhancement of the total SC order parameter $P_{\rm sc}$ is found at $dU/U\simeq 0.2$, at which $P_{\rm sc}$ is enhanced by a factor of $\sim 1.6$ over its value for
the degenerate case ($dU=0$), while the SC gap for the $x,y$ orbitals is enhanced by a factor of $\sim 3.5$ and
the critical temperature by a factor of $\sim 1.8$ (see Fig.\ref{fig:summary_results} for summary) !

We note that such a large enhancement cannot be obtained by an overall increase of $U/W$ for all of the orbitals, which would correspond
to an isotropic volume control of the material\cite{zadik_JTM_2015}.
Indeed, the chosen value $U/W=1$ is already quite close to $U/W\simeq 1.2$ at which $P_{\rm sc}$ is maximum,
and only a very small enhancement of $P_{\rm sc}$ can be obtained by increasing $U/W$ (see also Fig.\ref{fig:summary_results}).

Indeed, orbital differentiation is crucial to the effect, as seen from the orbitally-resolved gaps displayed in Fig.\ref{fig:imbalance_results}(a)
and from the orbital occupancies in Fig.\ref{fig:imbalance_results}(b).
Upon increasing $dU/U$, the SC gap associated with the two orbitals ($x,y$) having a smaller $U$ increases monotonically
up to $\sim0.07W$ at $dU/U=0.40$.
In contrast, $\Delta_{z}$ has a maximum and quickly decreases to a small value.
Correspondingly, the occupancy of the $x,y$ orbitals increases and that of the $z$-orbital decreases,
reaching limiting values at large $dU/U$.
These limiting values are easily understood qualitatively for a simplified interaction which would contain only density-density terms.
For $U_{zz}\gg U_{xx}=U_{yy}$,  a double occupancy is created in, say, the $x$-orbital,  while the
third electron is shared equally between $y$ and $z$ - an equally probable configuration having reversed roles of $x$ and $y$
which are identical by symmetry. (see Fig.\ref{fig:multiplets}(b))
Hence in this case $N_x=N_y=1/2(2+1/2)=1.25$ and $N_z=1/2$.
The limiting values we observe in our calculation for the full interaction (Eq.\ref{eq:Kanamori}) are somewhat reduced
from these ones by the pair-hopping term and charge fluctuations (Fig.\ref{fig:imbalance_results}, bottom panel).

These considerations suggest that the increased weight of configurations having a double occupancy in the
$x$ or $y$ orbital is crucial to the enhancement of the SC. Sec.~\ref{sec:mechanism} addresses the physical
mechanism underlying this effect in more details. We shall see there that the formation of a spin-singlet pair is indeed crucial,
but that an additional requirement is that these pairs can delocalize and become mobile through orbital fluctuations
and pair-hopping. This is indeed the case for $dU>0$, because the pair forms in either of the two degenerate,
symmetry-equivalent $x,y$ orbitals.

This also explains why a negative $dU$ is detrimental to SC.
In this case, formation of a pair is favored in the $z$-orbital,
as seen from the increase of $N_z$ in Fig.\ref{fig:imbalance_results}.
Since a single orbital is involved, the pair
is localized and cannot benefit from pair-hopping between orbitals.
Noteworthy is that SC is completely suppressed (at $dU/U\simeq -0.1$)
significantly before complete orbital polarization is reached
(corresponding to $N_z=2, N_{x,y}=1/2$, as indicated by dashed lines in Fig.\ref{fig:imbalance_results}(b)).

\begin{figure}[t]
\includegraphics[width=\columnwidth]{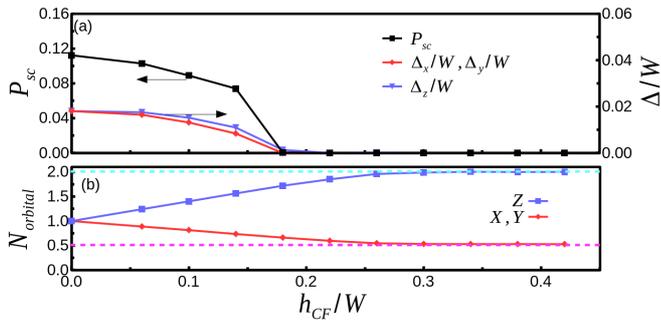}
\caption{(Color)
(a) SC order parameter $P_{\rm sc}$ and SC gaps $\Delta_{\alpha}/W$
as a function of the crystal-field, $h_{\rm CF}/W$ as defined in Eq.(\ref{eq:CF}).
(b) Orbital occupancies as a function of $h_{\rm CF}/W$.
Dashed lines indicate full orbital polarization into the $z$-orbital singlet configuration (see text).
\label{fig:xtalfield_results}
}
\end{figure}

\subsection{Crystal field}

Fig.~\ref{fig:xtalfield_results} displays our results in the presence of a crystal field splitting, Eq.~(\ref{eq:CF}).
The results are displayed for $\hcf>0$, since $\hcf<0$ is equivalent by symmetry (Appendix~\ref{sec:Negative crystal field}).
We see that this perturbation is detrimental to SC, in agreement with the recent findings of Ref.~\onlinecite{Hoshino_PRB_2016}.
Although the perturbation of $\hcf>0$ favors
the formation of a doubly occupied pair in the $z$-orbital, orbital fluctuations are suppressed and the pair cannot delocalize as for $dU<0$.
We note however that, in contrast to the $dU<0$ case, complete suppression of SC occurs only close to full orbital polarization
$N_z=2,N_{x,y}=0.5$.
Hence, one can say that SC is more robust in the presence of a crystal-field than in the presence of
an interaction imbalance with $dU<0$.
We discuss this difference in more details in Sec.~\ref{sec:mechanism}

\subsection{Summary and comparison of different perturbations}

\begin{figure}[t]
\includegraphics[width=\columnwidth]{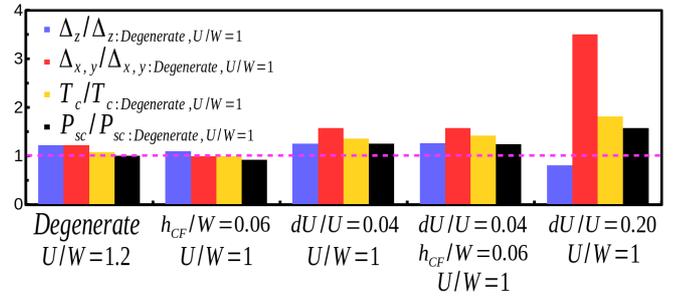}
\caption{(Color)
Summary of key results under various perturbations:
SC gaps, SC critical temperature $T_c$, and SC order parameter $P_{\rm sc}$, all normalized to their
corresponding values for $U/W=1$ in the degenerate case $dU=\hcf=0$.
\label{fig:summary_results}
}
\end{figure}

Fig.~\ref{fig:summary_results} provides a summary of some of our results for the key observables (SC order parameter,
gaps and $T_c$) and allows for a comparison between different cases, with reference to the $U/W=1$ degenerate case which
serves as a reference point.

\begin{itemize}
\item The results displayed for $U/W=1.2$ in the absence of any perturbation (degenerate case) demonstrate that only a very modest
enhancement of $T_c$ (by factor of $\sim1.1$) is obtained from a global increase of $U/W$ corresponding to a uniform compression
of the system, as stated above.

\item At the largest enhancement of the $P_{\rm sc}$, $dU/U=0.2$ (with $\hcf=0$), the remarkably large enhancements noted above are found,
by a factor $\sim 1.6$ for the total order parameter, $\sim 1.8$ for $T_c$ and $\sim 3.5$ for the gap $\Delta_{x,y}$.

\item We have also displayed results for a more modest imbalance $dU/U=0.04$. As explained in Sec.~\ref{sec:Tera-Hertz excitation}, this is our estimate
for a realistic value of $dU/U$ induced by THz pumping in the experiments of Mitrano et al.\cite{mitrano_K3C60_2016}.
Remarkably, a quite sizeable
enhancement of SC by a factor of $\sim1.35$ for $T_c$ is found there, despite the small value of $dU/U$. As also shown on Fig.~\ref{fig:summary_results},
this enhancement is not much affected by a small crystal field $\hcf/W\simeq 0.06$ (enhancement of $T_c$ by factor of $\sim1.41$
with respect to degenerate case)
which may also be induced by the THz pump (Sec.~\ref{sec:Tera-Hertz excitation}).

\end{itemize}

\section{Tera-Hertz excitation and $U$-Imbalance}
\label{sec:Tera-Hertz excitation}

In this section, we make contact with the experiment of Mitrano et al.\cite{mitrano_K3C60_2016} by estimating the
typical magnitude of the perturbations $dU/U$ and $\hcf/W$ induced by the mid-infrared optical pulse employed by
these authors.
It was proposed in Ref.~\onlinecite{mitrano_K3C60_2016} that this optical pulse excites the $T_{1u}(4)$ intra-molecular
vibrational mode of the C$_{60}$ molecule, and a typical value of $2.0\,{\AA}\sqrt{\mathrm{amu}}$ was quoted for
the normal coordinate excited amplitude of this mode. In turn, the perturbation modifies the electronic structure of the
$t_{1u}$ states. 

\begin{figure}[t]
\includegraphics[width=\columnwidth]{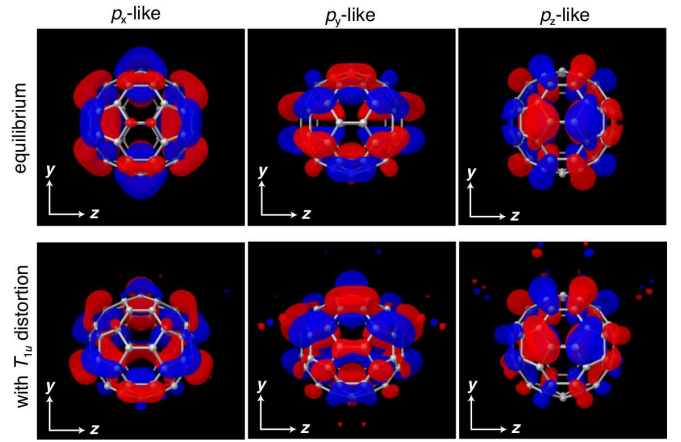}
\caption{(Color)
Top-row:  $t_{1u}$ Wannier functions for K$_3$C$_{60}$ in the equilibrium structure (top-row).
Bottom row: Wannier functions in the transient structure obtained by selectively exciting the $T_{1u}(4)$
phonon mode along the y-axis, at a maximum amplitude (corresponding to a $1/4$-cycle) of $2.0$ (${\AA}\sqrt{\mathrm{amu}}$)
- see Ref.\onlinecite{mitrano_K3C60_2016}.
Note that only the $p_z$-like orbital has a preserved nodal surface
in that case (corresponding to the $xy$ plane), leading to a larger value of $U$ for that orbital.
Only a single molecule is shown for visibility. The red (blue) color corresponds to positive (negative)
isosurfaces of the Wannier function.
In the transient structure
$p_{x}$-like and $p_{y}$-like orbitals (see text) do not have nodes but remain equivalent by symmetry.
(See Fig.\ref{fig:appendix_wannier})}
\label{fig:wannier}
\end{figure}

We have calculated the three Wannier functions associated with these electronic states in both
the unperturbed equilibrium structure, and in the perturbed structure corresponding to the
$T_{1u}(4)$ vibrational mode at the above quoted amplitude (i.e. for a quarter-cycle of the pulse).
The results are displayed in Fig.~\ref{fig:wannier}, for a polarization of the pulse conventionally chosen along the $y$-axis.
As expected, in the equilibrium structure (upper panel), we have three (degenerate) orbitals $p_x$, $p_y$ and $p_z$ having nodal surfaces
in the $yz$, $zx$ and $xy$ plane, respectively.
In the deformed structure (lower panel), the C-C bonds are distorted along the $x$- direction
with odd symmetry along the $y$-direction.
As a result, only the node in the $xy$ plane is preserved, corresponding to a $p_{z}$-like orbital
displayed on the lower right of Fig.\ref{fig:wannier}.
In contrast, as clear from Fig.\ref{fig:wannier}, the two other nodal planes ($zx$ and $yz$)
are absent in the perturbed structure, corresponding to two Wannier orbitals
that we still denote `$p_x$-like' and `$p_y$-like'.
While the inversion symmetry along the $y$-axis is broken by the perturbation,
inversion symmetry along the $x$- and $z$-axis is preserved, and the two orbitals
$p_x$ and $p_y$ are still degenerate.(See Fig.\ref{fig:appendix_wannier})

In order to evaluate the matrix elements of the screened Coulomb interaction in both the
equilibrium and perturbed structure, we have used the constrained-RPA (cRPA) ab-initio
method\cite{Ferdi_cRPA_2004}. In a nutshell, this method computes the effect of screening
in the random-phase approximation (RPA/GW), including all particle-hole excitations
except those for which both the initial and final states belong to the `target manifold' of $t_{1u}$
states retained in the low-energy many-body interaction Hamiltonian (Eq.\ref{eq:Kanamori}).
Matrix elements of this constrained screened interaction between the above Wannier states
are then calculated.
The results are displayed in Table~\ref{ImbalanceU}, calling for the following observations.

\begin{itemize}
\item
Overall, the effective intra-molecular repulsion $U_{\mathrm{eff}}$
(which can be approximately estimated as $U-V$ with $U$ the local component and $V$ the nearest-neighbour one,
see figure. 3 in Ref.~\onlinecite{nomurae_A3C60_sciadv})
is increased by the $T_{1u}(4)$ distortion for all orbitals, by a factor up to $\sim 1.2$.
This is because the $T_{1u}(4)$ distortion reduces the average distance
between electrons in the $t_{1u}$ states. Note however that, as discussed above, this overall increase of the
effective $U$ does not lead to a significant enhancement of SC.

\item The screened Coulomb interaction for the node-preserving $p_z$ orbital ($U_z=0.81$~eV, $U_z-V=0.58$~eV) is larger
than for the $p_x$, $p_y$ orbitals ($U_{x,y}=0.79$~eV, $U_{x,y}-V=0.56$~eV). This corresponds to a positive value of the
interaction imbalance of order $dU/U\simeq 0.04$ (considering the effective interaction $U-V$).
The physical reason for this is twofold. First, as shown on Fig.~\ref{fig:wannier}, the $p_z$ orbital in the distorted structure
displays an asymmetric charge distribution, with lobes of quite different sizes for $y>0$ as compared to $y<0$
because of the $y$-inversion symmetry-breaking. As a result,
there is an enhanced probability of having two electron closer to one another (for $y>0$ in the Fig.\ref{fig:wannier}) in this orbital.
In contrast, the suppression of the nodal surfaces for the $x,y$ orbitals leads to a more uniform charge distribution, so that
this asymmetric contraction is moderate for those orbitals.
Note that, in Ref.~\onlinecite{mitrano_K3C60_2016}, a calculation using a H\"{u}ckel model for the molecular orbitals
and simple electrostatics led to the opposite conclusion $dU<0$.
However, the reshaping of the molecular orbitals by the distortion (as well as screening) was not fully taken into account.

\item We have also calculated the value of the {\it electronic} (ferromagnetic) Hund's coupling in both the equilibrium
and distorted structure. A small increase is found ($0.008$~eV and $0.003$~eV for $x,y$ and $z$, respectively). The effect
of this on the effective inverted Hund's coupling would require an ab-initio evaluation of the change in the coupling to the
Jahn-Teller phonons, which we have not attempted here.
Hence, for simplicity, we kept $\Jinv$ constant in our study.

\end{itemize}

\begin{table*}[t]
\caption{Coulomb interaction matrix elements, as evaluated from cRPA,
in the equilibrium structure and in the presence of the $T_{1u}$ excitation.
The screened values correspond to the coupling constants acting on the low-energy $t_{1u}$ electronic states.
Values of the bare Coulomb interaction matrix elements without screening are also presented.
$U$ and $J_{\mathrm{H}}$ are on-site (local) couplings,
 while $V$ is the inter-site (nearest-neighbour) Coulomb interaction.
Calculations including both the $T_{1u}$ and $H_g$ distortions were also performed (not shown),
leading only to minor changes in the results for only $T_{1u}$ distortion case}.
\begin{ruledtabular}
\begin{tabular}{l  c  c  c  c  c}
           ~ & ~$U_{x,y}$ (eV)
             & ~$U_{z}$ (eV)
             & ~$J_{{\rm H}:xz,yz}$ (eV)
             & ~$J_{{\rm H}:xy}$ (eV)
             & ~$V$ (eV)~\\
\hline
Equilibrium (cRPA)    & 0.70 & 0.70 & 0.033 & 0.033 & 0.21\\
With $T_{1u}$ (cRPA)  & 0.79 & 0.81 & 0.041 & 0.036 & 0.23\\
Equilibrium (unscreened)    & 3.23 & 3.23 &  0.096 & 0.096 &  1.36    \\
With $T_{1u}$ (unscreened)  & 3.47 & 3.58 & 0.14 & 0.12 &  1.36 \\
\end{tabular}
\label{ImbalanceU}
\end{ruledtabular}
\end{table*}

So far, we have considered the direct effect of the pumped $T_{1u}(4)$ mode.
In Ref.~\onlinecite{mitrano_K3C60_2016}, it was suggested that this mode could also
induce a deformation of other structural modes through a non-linear coupling
(`non-linear phononics', see Refs.~\onlinecite{foerst_nlp_2011,alaska_nonlinear_2014}).
For example, a coupling of the form $Q^2_{T_{1u}}Q_{H_g}$ to a Raman-active $H_g$ mode
was considered, and its effect on the electronic structure of the $t_{1u}$ states
calculated~\cite{mitrano_K3C60_2016}.
The resulting crystal-field splitting in the perturbed structure (with both $T_{1u}(4)$ and $H_g$ perturbation)
was found to be $\hcf/W\simeq 0.06$. This is the motivation for also including this perturbation in the
model calculations reported above. We have found that a crystal-field of this magnitude does not
spoil the SC enhancement resulting from the $dU/U\simeq +0.04$ imbalance.


\section{Mechanism of superconductivity enhancement}
\label{sec:mechanism}

In this section, we discuss in some details the mechanism leading to SC enhancement or suppression by the two
perturbations considered in this article.

\subsection{Molecular eigenstates and histograms}
\label{sec:multiplets}

\begin{figure}[t]
\includegraphics[width=\columnwidth]{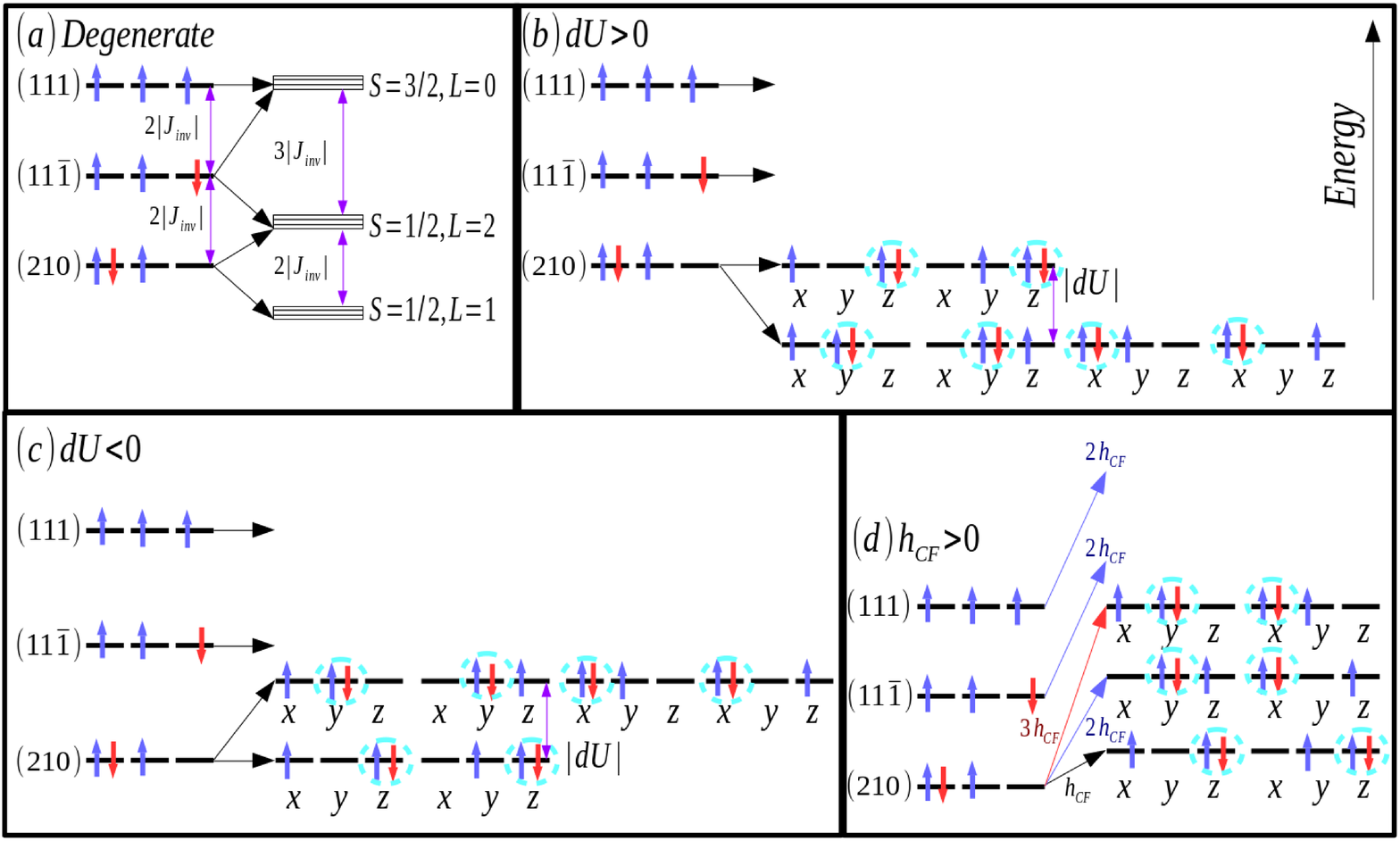}
\caption{(Color)
Multiplets of eigenstates for an isolated molecular site, for a half-filled
$t_{1u}$ shell occupied by three electrons.
(a): In the absence of any perturbation and when only density-density interactions are considered,
the (210), (11$\bar{1}$), and (111) states are eigenstates.
For the Kanamori hamiltonian (\ref{eq:Kanamori}) involving pair-hopping and spin-flip terms,
these states reorganize into ($S=3/2, L=0$), ($S=1/2, L=2$), and ($S=1/2, L=1$) multiplets, as
indicated by the arrows.
(b),(c),(d): Eigenstates for a density-density interaction hamiltonian,
in the presence of the perturbations $dU>0$, $dU<0$ and $\hcf$, respectively.
}
\label{fig:multiplets}
\end{figure}

First, we analyze the eigenstates (multiplets) of the interaction hamiltonian (\ref{eq:Kanamori}) for an isolated molecule,
and how they are modified by the perturbations. This analysis is summarized in Fig.~\ref{fig:multiplets}.

Consider first the inverted Hund's coupling interaction with density-density terms only (i.e omitting the pair-hopping and
spin-exchange terms contained in Eq.~(\ref{eq:Kanamori_dec})).
As displayed in panel (a) of Fig.~\ref{fig:multiplets}, the ground state is the (210) state made of a singlet pair in one of
the orbitals, a single electron in another orbital and an empty orbital. The excited states all have one electron in each
orbital, in either the (11$\bar{1}$) with one spin-flip or (111) spin-parallel configurations. These states are separated by
$2|\Jinv|$ and $4|\Jinv|$ from the ground-state, respectively.

For the full Kanamori hamiltonian (Eq.\ref{eq:Kanamori}) involving pair-hopping and spin-flip terms, these multiplets rearrange
into eigenstates of $S$ and $L$, as indicated by the arrows in panel~(a). The six-fold degenerate ground-state has
$(S=1/2,L=1)$: it can be viewed as a resonant superposition of the different (210) states with a singlet pair delocalized
between the different orbitals. The presence of a singlet pair in the ground-state and its coherent delocalization over
orbitals through the pair-hopping term is key to the SC found in the \scsc model.
This enhancement of SC by pair-tunneling between different orbitals was considered in Refs.~\onlinecite{suhl_bardeen_1959,kondo_superconductivity_1963},
and is sometimes referred to as the Suhl-Kondo mechanism.
Excited multiplets have $(S=1/2,L=2)$ and $(S=3/2,L=0)$ and are separated from the ground-state manifold by $2|\Jinv|$ and $5|\Jinv|$, respectively.

In panels (b,c,d) of Fig.~\ref{fig:multiplets}, we display the effect of the three perturbations $dU>0$, $dU<0$ and $\hcf$, respectively,
focusing for simplicity on a density-density interaction only.
The $dU>0$ perturbation (panel (b)) affects the (210) multiplets only. States with the spin-singlet
pair in the $z$-orbital are unaffected, while those with the pair in the $x$ or $y$ orbital are energetically stabilized,
by an amount $dU$. Hence, $dU>0$ leads to a relative stabilisation of (210) states with a pair in $x$ or $y$ orbital, as
compared to the (111) and (11$\bar{1}$) states.
This is a key ingredient for the promotion of SC by the $dU>0$ perturbation.

In contrast for $dU<0$ (panel (c)), the states with a pair in the $x$ or $y$ orbital are lifted up in energy, while
the molecular ground-state with a pair in the $z$-orbital is unaffected. This corresponds to a relative destabilization of
the (210) multiplet as compared to the (111) and  (11$\bar{1}$) ones.
%
Furthermore, because the pair is now localized in a single orbital, the pair-hopping term is ineffective.
Both factors lead to a suppression of SC.

\begin{figure}[t]
\includegraphics[width=\columnwidth]{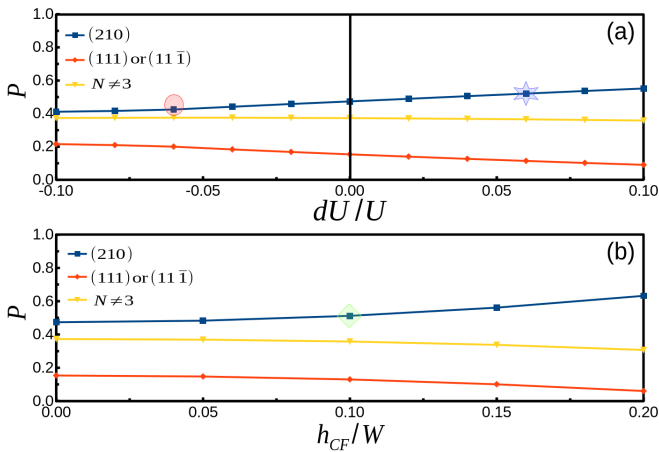}
\caption{(Color)
Statistical weights of the (210), (111) or (11$\bar{1}$) states in the presence
of (a) an interaction imbalance and (b) a crystal-field perturbation, as obtained from DMFT.
Datas indicated by circle (red), star (blue), and square (green) are analysed
in detail in Fig.\ref{fig:histograms2}.
}
\label{fig:histograms1}
\end{figure}

This analysis is based on the $t_{1u}$ many-body eigenstates of an isolated molecule, as described by (Eq.\ref{eq:Kanamori}).
In order to confirm its relevance to the full solid in the presence of inter-molecular hopping, we have calculated
the probability for a given state to be visited during the Monte-Carlo sampling performed in the full DMFT calculation~\cite{haule2007quantum}.
This is displayed in Fig.~\ref{fig:histograms1}. Panel (a) demonstrates that indeed the (210) states are stabilized by
the $dU>0$ perturbation, and destabilized for $dU<0$, relatively to the (111) or (11$\bar{1}$) states.

Fig.\ref{fig:multiplets}(d) displays the energy levels in the presence of a crystal-field with $h_{\rm CF}$.
The different components of the (210) ground-state manifold are lifted up in energy, by
$\hcf$, $2\hcf$ and $3\hcf$ depending on whether the $z$-orbital is doubly occupied, singly
occupied or empty, respectively. Excited states (111) and  (11$\bar{1}$) are lifted up by $2\hcf$.
As a result, the energy difference between the molecular (210) ground-state, which was
$2|\Jinv|$ in the absence of the perturbation, is now increased to $2|\Jinv|+\hcf$.
This corresponds to a relative stabilization of the (210) multiplet, as also demonstrated for the full
calculation by the statistical weights displayed in Fig.~\ref{fig:histograms1}(b).
We have seen, however, that SC is suppressed by the crystal-field perturbation. Hence, stabilization of
the (210) manifold is not a sufficient condition for the promotion of SC: pair delocalization through
orbital fluctuations is crucial, and we now analyze this in more details.

\subsection{Key role of orbital fluctuations}
\label{sec:fluctuations}

Orbital fluctuations within the multiplet of (210) states promote superconductivity.
The pair-hopping term (Eq.~\ref{eq:Kanamori_dec}) allows from the formation of an intra-molecular resonant
state with ($S=1/2,L=1$). In the degenerate case, the corresponding singlet pair can live
in any of the three orbitals, as illustrated in  Fig.\ref{fig:fluctuations}(a). The other panels in
this figure illustrate the fact that orbital fluctuations allowing the singlet pair to delocalize
between different orbitals are still possible for $dU>0$, but are hampered for both $dU<0$
and in the presence of a crystal field.

\begin{figure}[t]
\includegraphics[width=\columnwidth]{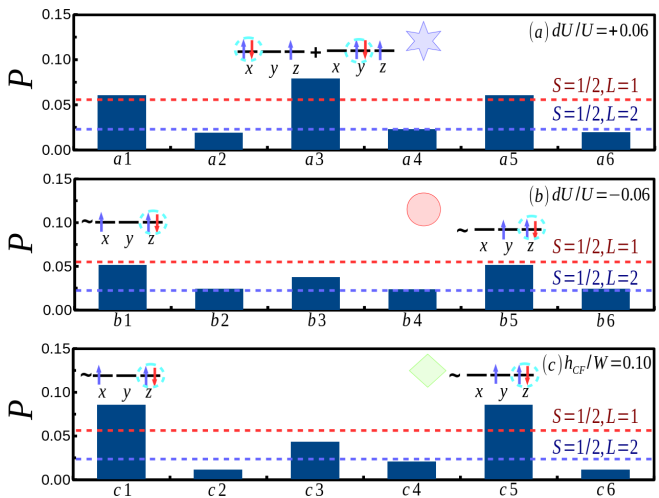}
\caption{(Color) Histogram of statistical weights  for (a) $dU/U=0.06$, (b) $dU/U=-0.06$, and (c) $h_{\rm CF}/W=0.10$.
The most dominant multiplets are plotted in each case.
See Fig.\ref{fig:histograms3} for the precise definition of eigenstates a1-a6, b1-b6, and c1-c6.
}
\label{fig:histograms2}
\end{figure}

Figure~\ref{fig:histograms2} presents a more in-depth analysis of the computed statistical weights of
states forming the (210) manifold, for three representative cases of $dU>0$, $dU<0$ and
crystal-field. This figure complements Fig.~\ref{fig:histograms1} in which only the global weight of
the (210) multiplet was presented. Here, we display the weights obtained by measuring
statistical weights of each eigenstate of the Hamiltonian in the (210) manifold
(detailed in Sec.~\ref{sec:Energy eigenstates}). We observe that:

\begin{figure}[t]
\includegraphics[width=\columnwidth]{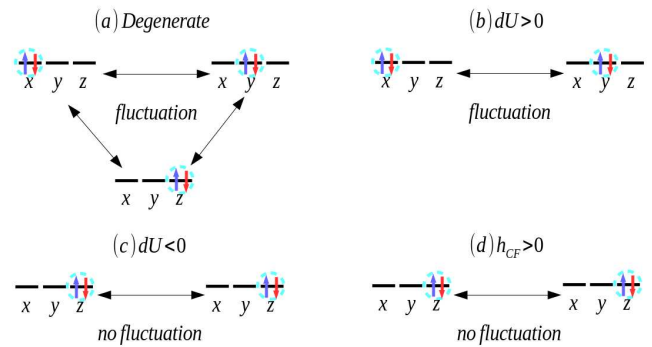}
\caption{(Color) This figures illustrates the key role of ground-state orbital fluctuations in promoting SC.
These fluctuations are present in the unperturbed degenerate case (a), are preserved for $dU>0$ (b),
while they are suppressed for $dU<0$ (c), or the presence of $\hcf$ (d).
Note that, for clarity, only the location of the pair in the ground-state is displayed in this figure
- the third lone electron has been omitted.
}
\label{fig:fluctuations}
\end{figure}

\begin{itemize}
\item For $dU>0$ (Fig.~\ref{fig:histograms2}(a)), the state which is a resonant superposition of a singlet pair
in the $x$ and $y$ orbitals (plus a lone electron in the $z$ orbital) has the largest statistical weight. Its weight is even larger
than that of the ($S=1/2,L=1$) states in the degenerate case. This fully confirms that the enhancement of SC observed
in this case originates from the stabilization of states involving a spin-singlet pair together while preserving orbital fluctuations.

\item For $dU<0$ (Fig.~\ref{fig:histograms2}(b)), the dominant states are those with a pair in the $z$ orbital only, hence suppressing orbital
fluctuations. Also the corresponding weight is reduced as compared to that of the ($S=1/2,L=1$) states in the degenerate case.
This confirms that the suppression of SC even when orbital polarization is incomplete originates from both the destabilization of singlet pair
configurations and in the suppression of orbital fluctuations.

\item In the presence of a crystal-field (Fig.~\ref{fig:histograms2}(c)), we have a mixed situation:
the weight of states having a spin-singlet pair is enhanced (by a factor $\sim 1.75$ in Fig.~\ref{fig:histograms2}(c)) as
compared to that of the ($S=1/2,L=1$) multiplet in the degenerate case. However, this increased weight corresponds to
states with a pair in the $z$ orbital only, for which orbital fluctuations are suppressed. As a result, the SC phase is resilient
as long as the orbital polarization is incomplete and is finally suppressed only when $\hcf/W$ is large enough to yield
 full orbital polarization and complete quenching of orbital fluctuations, as shown on Fig.\ref{fig:xtalfield_results}.

\end{itemize}

\subsection{Driving superconductivity into the strong-coupling regime}
\label{sec:scsc}

Finally, we demonstrate in this section that turning on the imbalance perturbation $dU>0$ increases
the effective pairing strength, i.e. that the system is driven into a more strongly coupled SC regime.

\begin{figure}[t]
\includegraphics[width=\columnwidth]{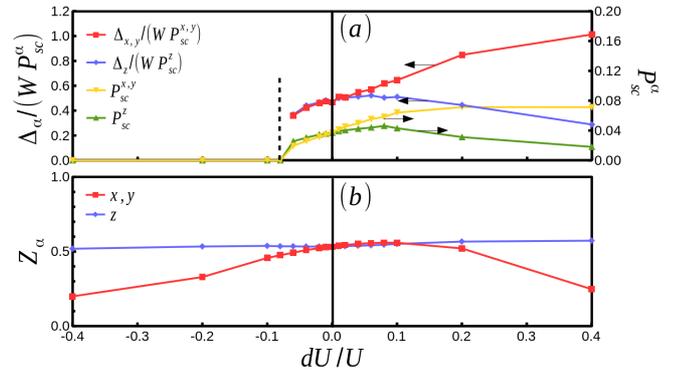}
\caption{
(Color) (a) Orbitally-resolved SC order parameter $P_{\rm sc}^\alpha$  and ratio $\Delta_\alpha/WP_{\rm sc}^\alpha$,
and (b) Orbitally-resolved quasiparticle weight $Z_\alpha$ as a function of interaction imbalance $dU/U$.
\label{fig:sc_ratio}
}
\end{figure}

To this aim, it is important to consider the different orbitals separately. In Fig.~\ref{fig:sc_ratio}(a), we display the
orbitally resolved SC order parameters $P_{\rm sc}^\alpha$ as a function of $dU/U$. As clear from this figure,
the action takes place in the $x$- and $y$-orbitals for which $P_{\rm sc}^{x,y}$ increases to large values with $dU/U$.
In contrast, after a small initial increase $P_{\rm sc}^z$ becomes quickly negligible. This corresponds to the dominant
pair-formation in a resonant state between the $x$- and $y$-orbital, as discussed before.

In order to reveal the effective SC coupling, we consider the dimensionless ratio $\Delta_\alpha/WP_{\rm sc}^\alpha$.
In a weak-coupling BCS superconductor, $\Delta=P_{\rm sc}\,V_{\mathrm{at}}$, where $V_{\mathrm{at}}$ is the effective
attractive interaction. Hence, this ratio equals $V_{\mathrm{at}}/W$ ($\ll 1$ in the BCS regime)
and is a good indicator of the effective dimensionless coupling.

This ratio is also displayed in Fig.~\ref{fig:sc_ratio}(a) as a function of $dU/U$. For $U/W=1,\Jinv/W=-0.04$ studied here, we see that it
takes a value $\sim 0.5$ in the degenerate case $dU=0$, indicating already a sizeable SC coupling. As $dU$ is increased,
this ratio becomes small for the $z$-orbital but increases rapidly for the $x,y$ orbitals, reaching a value of order unity at $dU/U=0.4$.
Hence the system is clearly driven into a strong-coupling SC regime by the interaction imbalance  perturbation.

It is interesting to compare and contrast this finding to the effect of uniformly increasing $U/W$ in the degenerate case:
$\Delta/WP_{\rm sc}$ is plotted {\it vs.} $U/W$ for this case in Fig.~\ref{fig:degenerate case} of Appendix.\ref{sec:Degenerate case}.
At small $U/W$ and $|\Jinv|/W$ (weak-coupling regime),
it takes a value close to $2|\Jinv|/W$, the pairing interaction predicted by BCS mean-field decoupling being $2|\Jinv|$ at $U=0$.
As $U/W$ is increased, it also increases to values of order unity. However, in this case, the Mott transition is reached at a critical
value $(U/W)_c\simeq 2$ at which the quasiparticle weight $Z$ vanishes. In contrast (Fig.~\ref{fig:sc_ratio}(b)) strong-coupling
superconductivity can be achieved by the imbalance perturbation without $Z_{x,y}$ becoming very small. Since
$Z$ enters the superfluid density, we conclude that the interaction imbalance perturbation permits a robust strongly-coupled superconductor without being limited by the proximity of the Mott transition.

\section{Conclusion and Perspectives}
\label{sec:conclusion}

Motivated by the recent experimental report of Mitrano {\it et al.}~\cite{mitrano_K3C60_2016},
we have investigated in this article possible theoretical mechanisms for enhanced superconductivity in \full fullerenes, based on \scsc model.

Our main finding is that an interaction imbalance corresponding to a smaller value of the
Coulomb matrix element for two of the orbitals in comparison to the third one, is indeed a perturbation
which efficiently enhances SC.
We have identified the physical mechanism responsible for this enhancement. 
It consists in a stabilisation of the intra-molecular states containing a singlet pair, while preserving the orbital
fluctuations associated with the pair-hopping interaction term. These fluctuations allow for the pair to be delocalized
coherently between two orbitals. Interaction imbalance was shown to bring SC further into the strong-coupling
regime. It provides a distinct and more efficient way to control and enhance SC in these compounds than the
volume-control corresponding to a uniform change of the bandwidth and interaction strength for all orbitals.

Furthermore, we have performed first-principle calculations of the changes of the electronic structure and
of the screened Coulomb matrix elements of \fullK, associated with the deformation due to
the pumped $T_{1u}$ intra-molecular mode. Our results demonstrate that indeed an interaction imbalance
with a favorable sign and magnitude for SC enhancement is induced by this deformation.

Our work calls for several future studies.
The amplitude of the modulation of the $T_{1u}$ mode quoted in Ref.\onlinecite{singla_molU_2015}
and used in the present work is but an order of magnitude estimate, and a more precise ab-initio
determination is desirable (as well as a direct experimental determination of the transient modulation
of the structure).
The light excitation may also induce other relevant changes in the material besides the
interaction imbalance considered here, such as more complex changes in the band-structure,
a modulation of the inter-orbital Coulomb matrix element and even a direct modulation of
the inverted effective Hund's coupling (due e.g. to an induced displacement and modulation of Jahn-Teller modes).
Inter-molecular phonons could also be an active part of pairing in the new excited structure~\cite{mitrano_K3C60_2016}.

Obviously however, the most pressing extension of our work is to perform a non-equilibrium study
in which the changes in electronic structure and orbital-dependent interaction strengths are time-dependent.
We note in this respect that, as noted in Ref.\onlinecite{singla_molU_2015}, the
dynamical selective excitation of a specific phonon mode leads to a
modulation of the Coulomb interaction at twice the frequency of the excited phonon.
In the present case, this corresponds to $\sim 0.4$~eV - a $10$~fs timescale,
100 times faster than the pico-second time scales over which the SC state is observed
in the experiments of Ref.~\cite{mitrano_K3C60_2016}.
Hence, the interaction modulation is in the antiadiabatic limit, providing some justification to the
equilibrium treatment performed in the present article.
Nonetheless, a full non-equilibrium treatment is in order and will be considered in future work.

\acknowledgements
We acknowledge useful discussions with Silke Biermann, Massimo Capone, Andrea Cavalleri, Stephen Clark, Dieter Jaksch, Giacomo Mazza
and Alaska Subedi (who also generously shared his calculations of the transient structures).
This work was supported by the European Research Council (ERC-319286 QMAC, ERC-617196 CORRELMAT, ERC-278472-MottMetals),
and by the Swiss National Science Foundation (NCCR MARVEL).

\appendix
\setcounter{secnumdepth}{0}
\section{APPENDIX}
\setcounter{secnumdepth}{3}
\subsection{Superconductivity in the degenerate case}
\label{sec:Degenerate case}

\begin{figure}[t]
\includegraphics[width=\columnwidth]{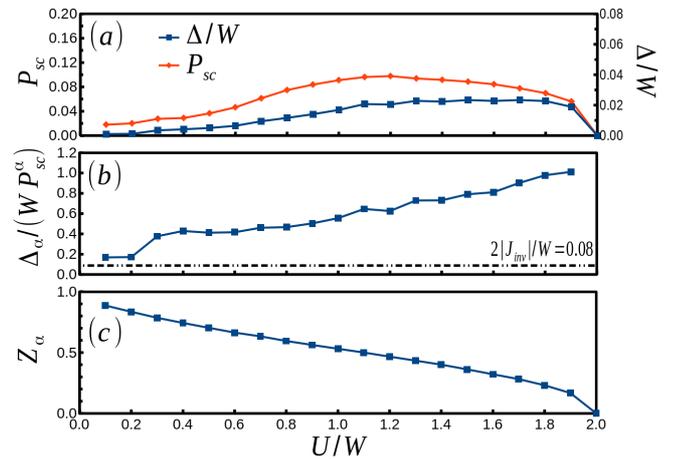}
\caption{(Color) Degenerate case (three equivalent orbitals).
(a) SC order parameter $P_{\rm sc}$, and gap $\Delta$, as a function of $U/W$/
(b) Ratio between SC gap and order parameter for each orbital $\Delta / WP_{\rm sc}^{\alpha}$,
(c) Quasiparticle weight $Z$ as function of $U/W$
The dashed-dotted line in (b) denotes the asymptotic value $2|\Jinv|/W$ of the attractive coupling which holds in the
weak-coupling BCS regime $U/W\rightarrow 0, |\Jinv|\ll W$ (See Sec.\ref{sec:scsc})
}
\label{fig:degenerate case}
\end{figure}

Figure~\ref{fig:degenerate case} displays $P_{\rm sc}$, $\Delta$,
$\Delta / P_{\rm sc}^{\alpha}$, and the quasiparticle spectral weight $Z$
as a function of $U/W$ in the degenerate case.
As discussed in the Sec.\ref{sec:scsc}, the enhancement of $U/W$
in the degenerate case drives the system into a strong-coupling SC
regime, as evidenced by the increasing ratio $\Delta / WP_{\rm sc}^{\alpha}$.
In contrast to Fig.\ref{fig:sc_ratio} however, the quasiparticle spectral weight
diminishes rapidly as shown in Fig.\ref{fig:degenerate case}(c), and the enhancement of SC is cutoff by entering the Mott phase.

\subsection{Equivalence of positive and negative crystal fields}
\label{sec:Negative crystal field}

\begin{figure}[t]
\includegraphics[width=\columnwidth]{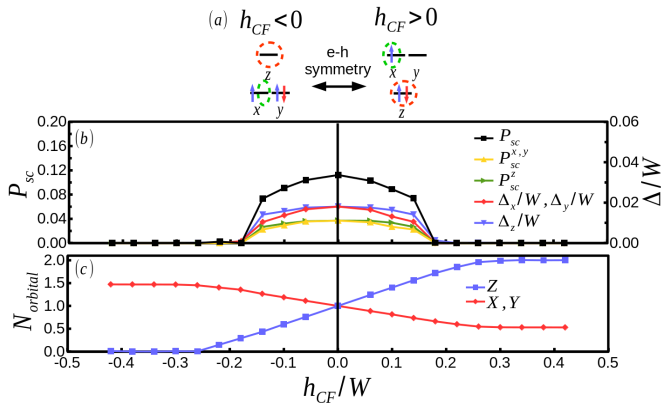}
\caption{(Color)
(a) Electron-hole symmetry
in the negative and positive $h_{\rm CF}$ cases.
(b) SC order parameter $P_{\rm sc}$,
orbitally-resolved pairing amplitude $P_{\rm sc}^{\alpha}$,
and SC gaps $\Delta_{\alpha}/W$
in the presence of perturbation of Eq.\ref{eq:CF}
for negative and positive $h_{\rm CF}$ values.
(c) Orbital occupancies for negative and positive $h_{\rm CF}$ values.
}
\label{fig:negativeCF}
\end{figure}

Figure~\ref{fig:negativeCF} illustrates the electron-hole symmetry
between $h_{\rm CF}>0$ and $h_{\rm CF}<0$ cases (see Ref.~\onlinecite{Hoshino_PRB_2016}).
As shown in Fig. \ref{fig:negativeCF}(a), a spin-singlet in the
electron picture for $h_{\rm CF}>0$ corresponds to a spin-singlet in the
hole picture for $h_{\rm CF}<0$ (red dotted circles).
A lone electron for $h_{\rm CF}>0$ corresponds to a lone hole for $h_{\rm CF}<0$
as shown in green dotted circles.

From this electron-hole correspondence,
$P_{\rm sc}$, $P^{\alpha}_{\rm sc}$, and $\Delta_{\alpha}$ exhibit
symmetric behavior with respect to $h_{\rm CF}=0$ axis as shown in Fig. \ref{fig:negativeCF}(b).
Also in Fig. \ref{fig:negativeCF}(c), the orbital occupancy for $h_{\rm CF}>0$ in the electron picture
 is symmetric with that for $h_{\rm CF}<0$ in the hole picture.
For positive $h_{\rm CF}$, full orbital polarization occurs around $h_{\rm CF}/W=0.3$
with orbital occupancies $N_{x,y}=0.5$ and $N_{z}=2$.
For negative $h_{\rm CF}$, full orbital polarization occurs around $h_{\rm CF}/W=-0.3$
with orbital occupancies $N_{x,y}=1.5$ and $N_{z}=0$.
In the hole picture, these occupancies correspond to 0.5 for $x$ and $y$, and 2 for $z$,
which is symmetric with positive $h_{\rm CF}/W=0.3$ case.

\subsection{Spectral function}
\label{sec:Spectral function}

\begin{figure}[t]
\includegraphics[width=\columnwidth]{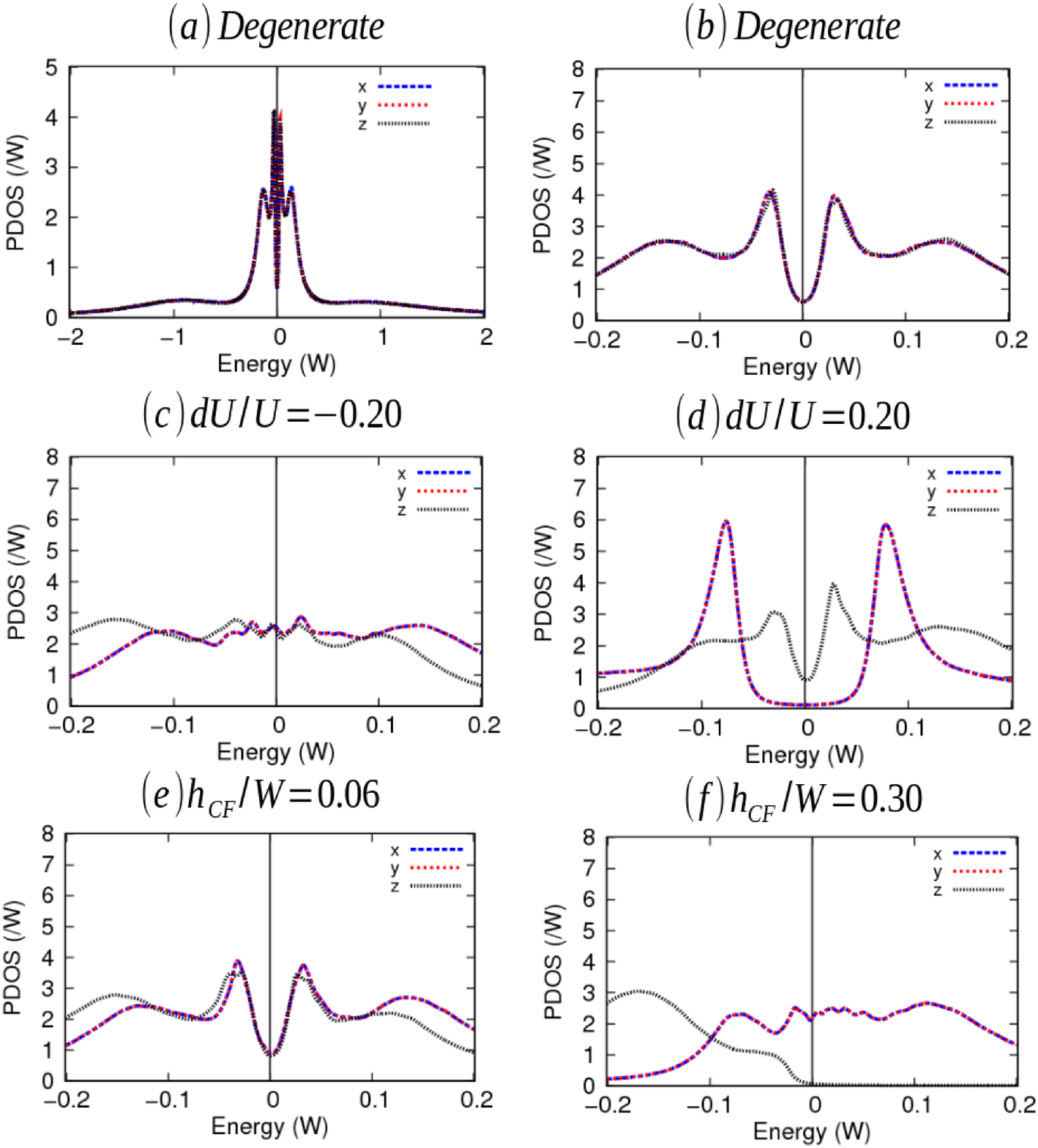}
\caption{(Color) Partial density of states (PDOS) for
orbital $x$, $y$, and $z$ in the
(a,b) degenerate, (c) $dU/U=-0.20$, (d) $dU/U=0.20$, (e) $h_{\rm CF}/W=0.06$,
and (f) $h_{\rm CF}/W=0.30$ cases.
Lorentzian smearing with the width of $0.004W$ is used in depicting PDOS.
}
\label{fig:density of states}
\end{figure}

Figure~\ref{fig:density of states} presents partial density of states (PDOS)
of orbitals $x$, $y$, and $z$ in the (a,b) unperturbed degenerate,
(c) $dU/U=-0.20$, (d) $dU/U=0.20$, (e) $h_{\rm CF}/W=0.06$,
and (f) $h_{\rm CF}/W=0.30$ cases.

In the unperturbed case Figs. \ref{fig:density of states}(a,b),
overall PDOS consists of the quasi-particle part, whose width is about $0.5W$, and
lower and upper Hubbard bands located at an energy scale of $\sim U = W$.
The width of the quasi-particle peak corresponds to the energy scale of Kondo temperature, and
well agrees with $Z W$ with the quasiparticle weight $Z$ ($\sim$ 0.5 for $U / W = 1$).
There exists a gap around the Fermi level due to the symmetry-breaking into SC phase.
The energy scale of the SC gap $\sim 0.04 W$
(see Fig.~\ref{fig:imbalance_results}, $U/W=0$) is smaller than that of Kondo temperature.
These two energy scales are also discussed in Ref.\onlinecite{capone_rmp_2009}

In the imbalanced Coulomb interaction with
$dU/U=-0.20$ case (Fig. \ref{fig:density of states}(c)), consistent with the results in Fig. \ref{fig:imbalance_results},
SC gaps are completely suppressed for all orbitals without full orbital polarization.
All $x$-, $y$-, and $z$-orbital PDOSs cross the Fermi level with up-shifting for $x$- and $y$-orbitals, and down-shifting for $z$-orbital, respectively.
In $dU/U=0.20$ case of Fig. \ref{fig:density of states}(d),
also consistent with the results in Fig. \ref{fig:imbalance_results},
SC gaps in $x$ and $y$-orbital, $\Delta_{x,y}$, are much larger than that of $z$-orbital, $\Delta_{z}$, i.e. SC gaps are anisotropic.

In the crystal-field case with $h_{\rm CF}/W=0.06$  (Fig. \ref{fig:density of states}(e)), PDOSs of $x$- and $y$-orbitals are shifted up,
and that of $z$-orbital is shifted down.
However, SC gaps for all orbitals are resilient, consistent with Fig.\ref{fig:xtalfield_results}.
The anisotropy in the size of SC gap is small.
For a large crystal-field, $h_{\rm CF}/W=0.30$ (Fig. \ref{fig:density of states}(f)), PDOS shows complete orbital polarization.
With this complete orbital polarization, SC gaps for all orbital are suppressed, consistent with Fig. \ref{fig:xtalfield_results}.

\subsection{{\it Ab initio} Calculations - Methods}
\label{sec:detail ab initio}

In the present study, we employ the {\sc quantum espresso} package~\cite{ESPRESSO_2009,QEspresso}
to perform {\it ab initio} calculations for fcc K$_3$C$_{60}$.
In the band structure calculation, we adopt the generalized-gradient approximation (GGA) with the Perdew-Burke-Ernzerhof parameterization~\cite{PBE_1996} for the exchange-correlation functional.
We prepare the Troullier-Martins norm-conserving pseudopotentials~\cite{TM_1991} in the Kleinman-Bylander representation~\cite{KM_1982} for the carbon and potassium atoms with the valence configurations
$(2s)^{2.0}(2p)^{2.0}$, and $(3p)^{6.0}$$(4s)^{0.0}$$(3d)^{0.0}$, respectively.
We take into account the nonlinear core correction~\cite{core_correction_1982} in the pseudopotential for the pottasium atom.
The calculation is done with 5$\times$5$\times$5 ${\mathbf k}$ mesh.
The cutoff energy for the wave functions is 40 Ry.
The calculations of the Coulomb interaction parameters are done within cRPA~\cite{Ferdi_cRPA_2004}.
The dielectric function is expanded with the plane waves with the energy cutoff of 7.5 Ry.
We employ 335 bands (129 occupied + 3 $t_{1u}$ bands + 203 unoccupied) to calculate the polarization function with exclude the transitions within $t_{1u}$ manifold.
The generalized tetrahedron method~\cite{doi:10.1143/JPSJ.72.777,PhysRevB.79.195110} was applied to perform the Brillouin-zone integral with respect to the wave vector.
We construct maximally localized Wannier functions~\cite{Marzari_wannier_1997,Souza_wannier_2001} from $t_{1u}$ manifold,
and the cRPA interaction parameters are given by the Wannier matrix elements of partially-screened Coulomb interactions.

The structures used in the calculation are same as Ref.~\onlinecite{mitrano_K3C60_2016}.
\begin{itemize}
\item fcc K$_3$C$_{60}$ with the lattice constant 13.89 \AA.
\item fcc K$_3$C$_{60}$ (a=13.89 \AA) + $T_{1u}(4)$ distortion (amplitude: 2.0 \AA$\sqrt{\rm amu}$)
\item fcc K$_3$C$_{60}$ (a=13.89 \AA) + $T_{1u}(4)+H_g(1)$ distortions (amplitudes: 2.0 and 1.5  \AA$\sqrt{\rm amu}$, respectively)
\end{itemize}
The obtained cRPA result for K$_3$C$_{60}$ in equilibrium structure is consistent with the estimate in Ref.~\onlinecite{nomura_C60_cRPA_2012}.

\subsection{Equivalence of $p_x$-like and $p_y$-like Wannier functions in Fig. \ref{fig:wannier} (bottom panel)}

\begin{figure}[t]
\includegraphics[width=\columnwidth]{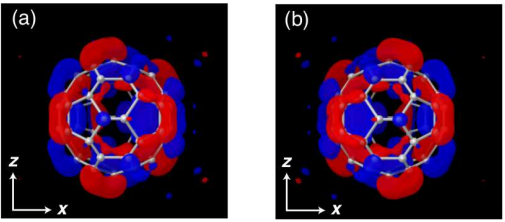}
\caption{(Color)
(a) Bottom-left and (b) bottom-center Wannier functions in Fig. \ref{fig:wannier}
viewed along $y$ axis.
Red (blue) color corresponds to positive (negative) isosurface of the wave function.
For visibility, only a single C$_{60}$ molecule in K$_3$C$_{60}$ solid is shown.
In (b), global phase factor $(-1)$ is multiplied to the Wannier function.
It is easy to see that these two Wannier functions are related by symmetry, and thus equivalent.
}
\label{fig:appendix_wannier}
\end{figure}

Figure~\ref{fig:appendix_wannier}(a) and \ref{fig:appendix_wannier}(b) show the bottom-left and bottom-center Wannier functions in Fig. \ref{fig:wannier}
viewed along $y$ axis.
Here $y$ axis is the direction of the $T_{1u}$ pumping.
We see that these two Wannier functions are related by symmetry, and thus are equivalent.

\subsection{Detailed composition of eigenstates in Fig.~\ref{fig:histograms2}}
\label{sec:Energy eigenstates}

\begin{figure}[t]
\includegraphics[width=\columnwidth]{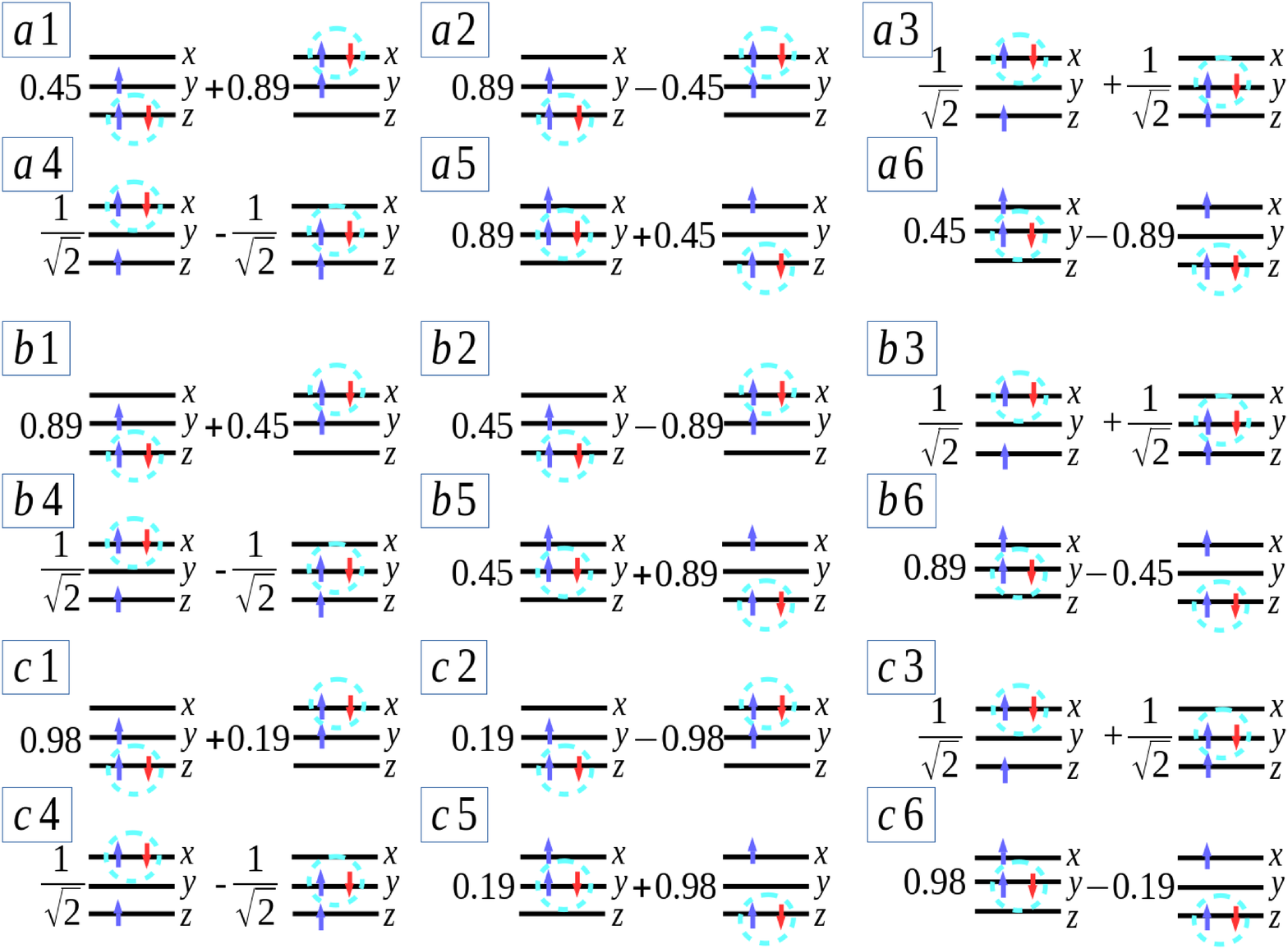}
\caption{(Color)
Energy eigenstates in the Fig.\ref{fig:histograms2},
a1-a6, b1-b6, and c1-c6.
}
\label{fig:histograms3}
\end{figure}

Figure~\ref{fig:histograms3} presents energy eigenstates in the
probability histogram of Fig.\ref{fig:histograms2}.

As shown in Fig. \ref{fig:histograms2},
for $dU/U=0.06$, the state which has the highest probability,
a3, shows symmetric resonance
between states with the spin-singlet in $x$- and $y$-orbitals.
Also noteworthy is that the probability of asymmetric resonant states between
the state with singlet in $z$-orbital and the state with a singlet in $x$ or $y$-orbital
as shown in a1 and a5 states, is larger than that of the ($S=1/2,L=1$) resonant state in the degenerate case.
This result implies that the stabilization of a $x$- and $y$-orbital singlet
also promotes the SC gap in the $z$-orbital, $\Delta_{z}$ for $dU/U=0.06$
which is consistent with Fig. \ref{fig:imbalance_results}(a).

In the case of $dU/U=-0.06$, in states which has the highest probability, b1 and b6,
the coefficient for the state with $z$-orbital spin-singlet (0.89)
is much larger than that of the state with $x$- or $y$-orbital singlet (0.45).
This suggests that the singlet electron-pair is localized mainly in $z$-orbital.
As a result, the orbital fluctuation is suppressed.

Also, in the case of $h_{\rm CF}/W=0.10$, in states with the highest probability,
c1 and c6, the coefficient for the state with $z$-orbital spin-singlet (0.98)
is much larger than that for state with $x$- or $y$-orbital spin-singlet (0.19).
Thus, the singlet-pair is localized in $z$-orbital, suppressing the orbital fluctuation.

\bibliography{refsK3C60}

\end{document}